\documentclass[a4paper,10pt]{article}
\pdfoutput=1

\usepackage[utf8]{inputenc}
\usepackage{amsmath}
\usepackage{graphicx}
\usepackage{caption}
\usepackage{subcaption}
\usepackage{amssymb}
\usepackage{color}
\usepackage[margin=1in]{geometry}
\usepackage{authblk}

\title{Boson Stars from Self-interacting Dark Matter}
\author[1]{Joshua Eby}
\author[2]{Chris Kouvaris}
\author[2]{Niklas Grønlund Nielsen}
\author[1]{L.C.R. Wijewardhana}
\affil[1]{\emph{University of Cincinnati, Dept. of Physics \\ 
  Cincinnati, OH 45221 USA}}
\affil[2]{\emph{CP$^3$-Origins \\
  University of Southern Denmark, Campusvej 55, DK-5230, Odense M, Denmark}}

\begin{document}
\maketitle

 \begin{abstract}
 We study the possibility that self-interacting bosonic dark matter forms star-like objects. We study both the case of attractive and repulsive self-interactions, and we focus particularly in the parameter phase space where self-interactions can solve well standing problems of the collisionless dark matter paradigm. We find the mass radius relations for these dark matter bosonic stars, their density profile as well as the maximum mass they can support. 
 \\[.1cm]
 {\footnotesize  \it Preprint: CP3-Origins-2015-044 DNRF90}
\end{abstract}

\section{Introduction}
Bosonic degrees of freedom arise generically and naturally in theories of fundamental physics, both in the Standard Model and beyond.  The Higgs boson is of paramount importance, being the only fundamental scalar in the Standard Model \cite{Higgs:1964pj,Chatrchyan:2012xdj}, but many other scalar degrees of freedom have been proposed to extend particle physics to high energy scales.  These include (among many others) the axion of QCD \cite{Peccei:1977hh} or the scalar which drives the expansion of the universe in quintessence models \cite{Ratra:1987rm}.  

These bosonic particles often make good Dark Matter (DM) candidates as well.  One reason for this is that unlike the Higgs, many of these new scalars would be stable or long-lived enough that they could coalesce into DM halos which constitute the seeds of galaxy formation.  Unlike the usual collisionless cold DM picture, however, we are interested in the scenario where large collections of these bosons form bound states of macroscopic size due to their self-gravitation (and self-interaction generically).  For this picture to be consistent, the scalars are taken to be sufficiently cold so that they may coalesce into a Bose-Einstein Condensate (BEC) state, and can thus be described by a single condensate wavefunction.  These wavefunctions can indeed encompass an astrophysically large volume of space and have thus been termed ``boson stars'' \cite{Colpi:1986ye}.

It was shown many years ago that objects of this type are allowed by the equations of motion, first by Kaup \cite{Kaup:1968zz} and subsequently by Ruffini and Bonazzola \cite{Ruffini:1969qy} in non-interacting systems.  They found a maximum mass for boson stars of the form $M_\text{max} \approx 0.633 M_\text{P}^2/m$, where $M_\text{P}=1.22\times10^{19}$ GeV is the Planck mass and $m$ is the mass of the individual bosons.  (This is very different from the analogous limit for fermionic stars, termed the Chandrasekhar limit, which scales as $M_\text{P}^3/m^2$).  Later, it was shown by Colpi et al. \cite{Colpi:1986ye} that self interactions in these systems can cause significant phenomenological changes.  In particular, they examined systems with repulsive self-interactions, and show that the upper limit on the mass is $M_\text{max}\approx 0.02\sqrt{\lambda}M_\text{P}^3/m^2$, where $\lambda$ is a dimensionless $\phi^4$ coupling.\footnote{Note that the Colpi et al. result does not reduce to the Kaup bound as $\lambda\rightarrow 0$ because the former is derived by rescaling the equations of motion and dropping higher-order terms in the strong coupling limit, as we see in Section \ref{BosonSec}.}    
This extra factor of $M_\text{P}/m$ as compared to the noninteracting case makes it more plausible that boson stars can have masses even larger than a solar mass.  A different method of constraining the boson star parameter space, which fits the coupling strength using data from galaxy and galaxy cluster sizes, has been considered in \cite{Souza1,Souza2}.

The situation for attractive self-interactions is slightly more complex.  The simplest case involves a self-interaction of the form $\lambda \phi^4$, where $\lambda<0$ for attractive interactions.  If this were the highest-order term in the potential, then it would not be bounded below, and so one typically stabilizes it by the addition of a positive $\phi^6$ term.  We will assume that the contribution of such higher-order terms is negligible phenomenologically (we address the validity of that assumption in Section \ref{3.2}).  Furthermore, in this scenario the typical sizes of gravitationally bound BEC states is significantly smaller than the repulsive or non-interacting cases.  This is because the only force supporting the condensate against collapse comes from the uncertainty principle. Gravity and attractive self-interactions tend to shrink the condensate.  We will see in Section \ref{BosonSec} that the maximum mass for an attractive condensate scales as $M_\text{max} \sim M_\text{P}/\sqrt{|\lambda|}$. This result was originally derived using an approximate analytical method \cite{Chavanis:2011zi}, and was later confirmed by a precise numerical calculation \cite{Chavanis:2011-2}.

DM self-interactions have already been  proposed and studied in different contexts~\cite{Spergel:1999mh,Wandelt:2000ad,Faraggi:2000pv,Mohapatra:2001sx,Kusenko:2001vu,Loeb:2010gj,Kouvaris:2011gb,
Rocha:2012jg,Peter:2012jh,Vogelsberger:2012sa,Zavala:2012us,Tulin:2013teo,Kaplinghat:2013xca,Kaplinghat:2013yxa,Cline:2013pca,
Cline:2013zca,Petraki:2014uza,Buckley:2014hja,Boddy:2014yra,Schutz:2014nka}. One of the main reasons why DM self-interactions can play an important role is due to  the increasing tension between numerical simulations of collisionless cold DM and astrophysical observations, the resolution of which (for the moment) is unknown.  The first discrepancy, known as the ``cusp-core problem'', is related to the fact that dwarf galaxies are observed to have flat density profiles in their central regions \cite{Moore:1994yx,Flores:1994gz}, while N-body simulations predict cuspy profiles for collisionless DM \cite{Navarro:1996gj}.  Second, the number of satellite galaxies in the Milkly Way is far fewer than the number predicted in simulations \cite{Klypin:1999uc,Moore:1999nt,Kauffmann:1993gv,Liu:2010tn,Tollerud:2011wt,Strigari:2011ps}.  Last is the so-called ``too big to fail'' problem: simulations predict dwarf galaxies in a mass range that we have not observed, but which are too large to have not yet produced stars \cite{BoylanKolchin:2011de}.  

The solution of these problems is currently unknown, but a particularly well-motivated idea involves self-interacting DM (SIDM).  Simulations including such interactions suggest that they have the effect of smoothing out cuspy density profiles, and could solve the other problems of collisionless DM as well \cite{Dave:2000ar, Vogelsberger:2013eka, Rocha:2012jg}.  These simulations prefer a self-interaction cross section of $0.1$ cm$^2$/g $\lesssim \sigma/m \lesssim 10$ cm$^2$/g.  There are, however, upper bounds on $\sigma/m$ from a number of sources, including the preservation of ellipticity of spiral galaxies \cite{Feng:2009mn,Feng:2009hw}.  The allowed parameter space from these constraints nonetheless intersects the range of cross sections which can resolve the small-scale issues of collisionless DM, in the range $0.1$ cm$^2$/g $\lesssim \sigma/m \lesssim 1$ cm$^2$/g. 

Self-gravitation and additionally extra self-interactions among DM particles can lead in some cases to the collapse of part of the DM population into formation of dark stars. The idea of DM forming star-like compact objects is not new. Dark stars that consist of annihilating DM might have existed in the early universe~\cite{Spolyar:2007qv,Freese:2008hb,Freese:2008wh}. Dark stars have been also studied in the context of hybrid compact stars made of baryonic and DM~\cite{Leung:2011zz,Leung:2013pra,Tolos:2015qra,Mukhopadhyay:2015xhs} as well as in the context of mirror DM~\cite{Khlopov:1989fj,Silagadze:1995tr,Foot:1999ex,Foot:2000iu}. Additionally some of the authors of the current paper studied the possibility of dark star formation from asymmetric fermionic DM that exhibits Yukawa type  self-interactions that can alleviate the problems of the collisionless cold DM paradigm~\cite{Kouvaris:2015rea}. Unlike the dark stars of annihilating DM, asymmetric dark stars can be stable and observable today. \cite{Kouvaris:2015rea} displays the parameter space where it is possible to observe such dark stars, providing mass radius relations, corresponding Chandrasekhar mass limits and density profiles.  Self-interactions in dark stars have also been considered in \cite{Schaffner-Bielich_Fermions} for fermionic particles, as well as in \cite{Schaffner-Bielich_Bosons} for bosonic ones.
 
In this paper we examine the dark stars composed of asymmetric self-interacting bosonic DM. The study is fundamentally different from that of ~\cite{Kouvaris:2015rea} because unlike the case of fermionic DM where the stability of the star is achieved by equilibrium between the Fermi pressure and gravitation, bosonic DM does not have a Fermi surface. They form a BEC in the ground state and it is the uncertainty principle that keeps the star from collapsing. We are going to demonstrate how DM self-interactions affect the mass radius relation, the density profile and the maximum mass of these DM bosonic stars in the context of the self-interactions that reconcile cold DM with the observational findings.

Note that we set $\hbar=c=1$ in what follows.

\section{SIDM parameter space} \label{AstroSec}
As we mentioned above, galactic scale $N$-body simulations of cold, non-interacting DM indicate that the central regions of galaxies should have a ``cuspy'' density profile, contrary to the cored profiles one observes.  This, along with the ``missing satellites'' and ``too big to fail'' problems, has led some to question the non-interacting DM paradigm.  While some believe that the inclusion of baryonic physics could alleviate these issues \cite{Oh:2010mc,Brook:2011nz,Pontzen:2011ty,Governato:2012fa}, it remains an open question.  On the other hand, the inclusion of self-interactions in the DM sector could resolve these issues without creating tension with other astrophysical constraints. These two conditions can be simultaneously satisfied if the cross section per unit mass for DM satisfies
\begin{equation} \label{csConstraint}
 0.1 \frac{\text{cm}^2}{\text{g}} \lesssim \frac{\sigma}{m} \lesssim 1 \frac{\text{cm}^2}{\text{g}}.
\end{equation}
Assuming a velocity independent cross section, \cite{Rocha:2012jg} found that $\sigma/m = 1$ cm$^2$/g tends to over-flatten dwarf galaxy cores and that it is marginally consistent with ellipticity constraints of the Milky Way. On the other hand a value of $0.1$ cm$^2$/g satisfies all constraints and flattens dwarf galaxy cores sufficiently.
Let us  consider a potential of the form 
\begin{equation}
 V(\phi) = \frac{m^2}{2}\phi^2 + \frac{\lambda}{4!} \phi^4. 
\end{equation}
Note that $\lambda>0$ ($\lambda<0$) signifies a repulsive (attractive) interaction. The resulting DM-DM scattering cross-section is
\begin{equation}
 \sigma = \frac{\lambda^2}{64\pi m^2}
 \label{eq:sigmaSC}
\end{equation}
at tree level.  Plugging this into Eq. (\ref{csConstraint}), we get the constraint
\begin{equation}\label{bounds}
 \Big(\frac{m}{1\text{ MeV}}\Big)^{3/2} < \frac{|\lambda|}{10^{-3}} 
			< 3 \Big(\frac{m}{1\text{ MeV}}\Big)^{3/2}.
\end{equation}
This matches the results of \cite{AmaroSeoane:2010qx}.  For perturbativity, we should restrict $\lambda \lesssim 4\pi$, which would imply that our results are valid only for $m\lesssim 100$ MeV.  In this mass range, it is plausible that these DM particles coalese into boson stars at some point in early cosmology. 

If a large fraction of DM is contained inside boson stars, the derived parameter space may be significantly altered \cite{Enqvist:2001jd}, since boson star-DM interactions and boson star self-interactions may become significant. We will however assume that boson stars are rather scarce and the DM self-interactions are dominated by DM-DM scattering. 

\subsection{DM scattering with boson stars}
 To quantify how scarce boson stars have to be within this approximation, we assume that boson stars have a characteristic radius $R$, mass $M$ and number density $n_\text{BS}$. The mass, number density and self-interaction cross section of free DM is taken to be $m$, $n$ and $\sigma$. The mean free path a DM particle travels before hitting another DM particle or a boson star will be $\lambda_{DM}= (n \sigma)^{-1} $ and $\lambda_\text{BS} \sim (n_\text{BS} \pi R^2)^{-1} $, respectively. Scattering with boson stars has to be much rarer than with other free DM in our approximation. Therefore we require $\lambda_\text{DM} \ll \lambda_\text{BS}$. For the DM density we use the typical value of the solar system, i.e. $\rho_\text{DM} = M n_\text{BS} + m n \approx 0.3 \text{GeV/cm}^3$. These requirements lead to the following condition
\begin{equation}
n_\text{BS} \ll \frac{\sigma \rho_\text{DM}}{m\pi R^2+M\sigma}.
\label{eq:BSnumberdensity}
\end{equation}
Taking self-interactions to be that of Eq.~(\ref{eq:sigmaSC}), and the boson star radius to be comparable to the minimum radius (which scales the same for both signs of interaction)
$R \sim \sqrt{|\lambda|} M_\text{P}/m^2$ (see Eq.~(\ref{R_Att})),
Eq.~(\ref{eq:BSnumberdensity}) becomes
\begin{equation}
n_\text{BS} \ll \frac{\rho_\text{DM}}{64 \pi^2\tfrac{M_\text{P}^2}{|\lambda| m} + M}.
\label{eq:BSnumberdensity2}
\end{equation}
The maximum mass of a boson star with non-negligible attractive interactions is $\sim M_\text{P}/\sqrt{|\lambda|}$. Since this scaling is only proportional to a single power of $M_\text{P}$, the first term in the denominator of Eq.~(\ref{eq:BSnumberdensity2}) tends to dominate. We obtain in the attractive scenario
\begin{equation}
n_\text{BS}^\text{att} \ll \frac{|\lambda| m\rho_\text{DM}}{64 \pi^2M_\text{P}^2 } \approx 2\times 10^{-5} |\lambda| \frac{m}{\text{MeV}} \text{AU}^{-3},
\end{equation}
where AU is an astronomical unit. The minimum mean distance between attractive boson stars can therefore within this approximation be $(n_\text{BS}^\text{att})^{-1/3} \approx 40 (|\lambda| m/\text{MeV})^{-1/3} \text{AU}$.
In the scenario with repulsive interactions the maximum mass scales as $\sqrt{\lambda} M_\text{P}^3/m^2$. Therefore the second term in the denominator of Eq.~(\ref{eq:BSnumberdensity2}) dominates. The number density must satisfy
\begin{equation}
n_\text{BS}^\text{rep} \ll \frac{m^2\rho_\text{DM}}{\sqrt{\lambda} M_\text{P}^3} \approx 9\times 10^{-9} \lambda^{-1/2}\left( \frac{m}{\text{MeV}}\right)^2 \text{pc}
^{-3}.
\end{equation}
The minimum mean distance between repulsive boson stars which leaves our approximation valid can at most be $(n_\text{BS}^\text{rep})^{-1/3} \approx 5 \times 10^2 \lambda^{2/3} (m/\text{MeV})^{-2/3} \text{pc}$.

\section{Bosonic Dark Matter} \label{BosonSec}
An important property of light scalar particles that has been examined extensively in the literature \cite{Matos:2008ag,Suarez:2011yf} is that large collections (particle number $N\gg1$) can transition to a BEC phase at relatively high temperature, as compared to terrestrial experiments with cold atoms.  The critical temperature for condensate occurs when the de Broglie wavelength is equal to the average interparticle distance, $\lambda_\text{dB}=[\zeta(3/2)/n]^{1/3}$, where $n$ is the average number density of the particles and $\zeta(x)$ is the Riemann Zeta function.  This implies a critical temperature for transition to the BEC phase of the form
\begin{equation}
 k T_\text{c} = \frac{2\pi}{m}\Big(\frac{n}{\zeta(3/2)}\Big)^{2/3}.
\end{equation}
In this paper, we will assume that all relevant scalar field particles are condensed, i.e. that the  system is in its ground state, a perfect BEC.  The effect of thermal excitations is examined in \cite{Harko:2011dz} and they are expected to be negligible as long as $T<T_\text{c}$ is satisfied.

\subsection{Non-Interacting Case}
It is instructive to begin with the case of boson stars bound only by gravity, first analyzed in \cite{Kaup:1968zz}.  In this seminal work, Kaup considers the free field theory of a complex scalar in a spacetime background curved by self-gravity.  The equations of motion\footnote{The non-interacting equations of motion are equivalent to Eq.s (\ref{EKG_Colpi}) and (\ref{Metric_Colpi}) in the limit $\Lambda\rightarrow0$.} were solved numerically.  The maximum mass of these solutions was found to be $M_{max}\approx 0.633 M_P^2/m$, the oft-quoted Kaup limit for non-interacting boson stars.  This value was later confirmed by Ruffini and Bonazzola \cite{Ruffini:1969qy}, who used a slightly different method by taking expectation values of the equations of motion in an $N$-particle quantum state.

Interacting field theories are more complex.  In particular, for cross sections satisfying Eq. (\ref{csConstraint}), the phenomenology of repulsive and attractive interactions are very different, and accordingly, the methods required to analyze them are different as well.  We outline the relevant methods in the sections below.

\subsection{Repulsive Interactions}
If the self-interaction is repulsive, we can make use of the result of Colpi et al. \cite{Colpi:1986ye}. Like Kaup, their method begins with the relativistic equations of motion for a boson star, the coupled Einstein and Klein-Gordon equations, but including a self-interaction term represented by $\Lambda$:
\begin{align} \label{EKG_Colpi}
 \frac{A'}{A^2 x} + \frac{1}{x^2}\left(1-\frac{1}{A}\right)
	&= \left(\frac{\Omega^2}{B}+1\right)\sigma^2 + \frac{\Lambda}{2}\sigma^4
		    + \frac{(\sigma')^2}{A} \nonumber \\
 \frac{B'}{B^2 x} + \frac{1}{x^2}\left(1-\frac{1}{A}\right)
	&= \left(\frac{\Omega^2}{B}+1\right)\sigma^2 - \frac{\Lambda}{2}\sigma^4
		    + \frac{(\sigma')^2}{A} \nonumber \\
 \sigma'' + \left(\frac{2}{x} + \frac{B'}{2B} - \frac{A'}{2A}\right)&\sigma'
	+ A \left[\left(\frac{\Omega^2}{B}-1\right)\sigma 
		    - \Lambda\sigma^3\right] = 0,
\end{align}
where the rescaled variables are $x=mr$, $\sigma=\sqrt{4\pi G}\Phi$ ($\Phi$ the scalar field), $\Omega = \omega/m$ ($\omega$ the particle energy), and $\Lambda=\lambda M_\text{P}^2/(4\pi m^2)$.  In addition to the scalar field itself, $A(r)$ and $B(r)$ must be solved for; these represent the deviations from the flat metric due to the self-gravity of the condensate,
\begin{equation} \label{Metric_Colpi}
 ds^2 = -B(r) dt^2 + A(r) dr^2 + r^2 d\Omega^2.
\end{equation}
In practice, one can trade the metric function $A(r)$ for the mass $\mathcal{M}(x)$ by the relation $A(x)=[1-2\mathcal{M}(x)/x]^{-1}$. In the limit that the interactions are strong (precisely, $\Lambda \gg 1$), the system can be simplified significantly, as one can perform a further rescaling of the equations: $\sigma_* = \sigma\Lambda^{1/2}$, $x_*=x\Lambda^{-1/2}$, and $\mathcal{M}_*=\mathcal{M}\Lambda^{-1/2}$.  The relevant parameters of Section \ref{AstroSec} suggest a value of $\Lambda = \mathcal{O}(10^{40})$ or higher, so it is completely safe to neglect terms proportional to $\Lambda^{-1}$. In this limit the equations simplify to
\begin{align}
 \sigma_* &= \sqrt{\frac{\Omega^2}{B}-1} \nonumber \\
 \mathcal{M}_*' &= 4\pi x_*^2\rho_* \nonumber \\
 \frac{B'}{Bx_*}\left(1-\frac{2\mathcal{M}_*}{x_*}\right)
	      &-\frac{2\mathcal{M}_*}{x_*^3}=8\pi p_*,
\end{align}
where the pressure $p_*$ and density $\rho_*$ are given by
\begin{align}
 \rho_* &= \frac{1}{16\pi}\left(\frac{3\Omega^2}{B}+1 \right)
	      \left(\frac{\Omega^2}{B}-1 \right) \nonumber \\
 p_* &= \frac{1}{16\pi}\left(\frac{\Omega^2}{B}-1 \right)^2 .
 \label{Eq:EoS}
\end{align}
In this limit, the equations do not depend on $\Lambda$, and one finds numerically that there is a maximum (dimensionless) mass $\mathcal{M}_{*\text{max}}\approx 0.22$.  Restoring the appropriate dimensions, one finds
\begin{equation} \label{mc_att}
 M < M_\text{max}^\text{rep} = 0.22\sqrt{\frac{\lambda}{4\pi}}\frac{M_\text{P}^3}{m^2}.
\end{equation}
This bound on the mass of repulsive boson stars was confirmed very precisely using a hydrodynamic approach as well \cite{Chavanis-GR}.

\begin{figure}[htc]
\centering
\begin{minipage}{0.46\textwidth}
\centering
 \includegraphics[width=.9\textwidth]{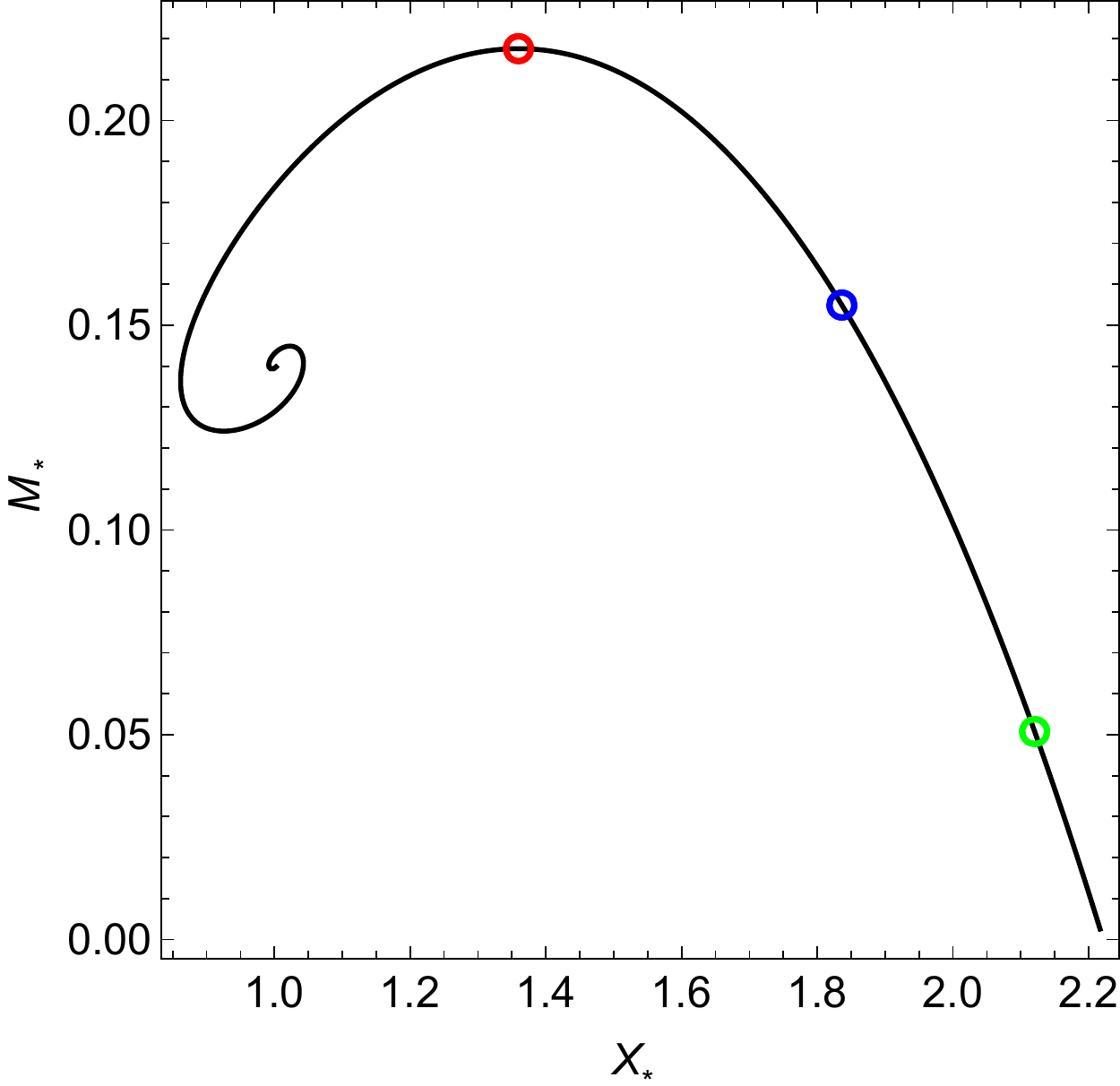}
\caption{The mass-radius relation for a boson star with strong repulsive coupling. The 3 circles correspond to the density profiles in Figure \ref{Colpi_Density_plot}.  The dimensionless variables in the plot are defined in terms of the dimensionful ones as $M_*= m M^2\Lambda^{-1/2}/M_\text{P}$ and $X_*= mR\Lambda^{-1/2}$.}
\label{Colpi_MassRad_plot}
\end{minipage}\hfill
\begin{minipage}{0.46\textwidth}
\centering
 \includegraphics[width=.9\textwidth]{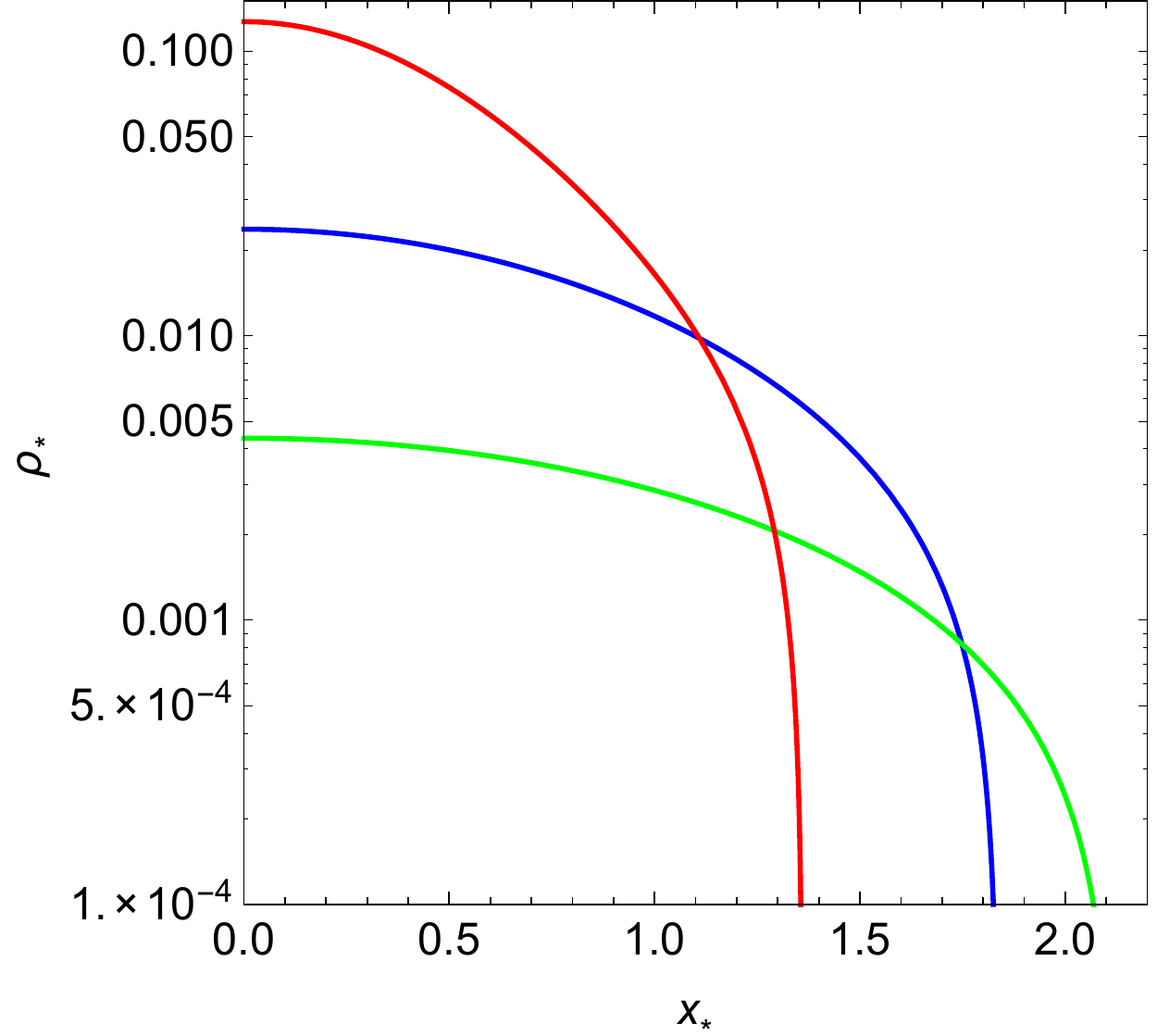}
\caption{Three examples of density profiles in the case of repulsive interactions. The red profile corresponds to the profile of the maximum mass equilibrium, while the blue and green are taken on the stable branch of equilibria.  The dimensionless variables in the plot are defined in terms of the dimensionful ones as $\rho_*$ defined in Eq. (\ref{Eq:EoS}) and $x_*=mr\Lambda^{-1/2}$.}
\label{Colpi_Density_plot}
\end{minipage}
\end{figure}

Figures \ref{Colpi_MassRad_plot} and \ref{Colpi_Density_plot} show the mass-radius relation and selected density profiles, respectively. The branch to the left of the peak in Figure \ref{Colpi_MassRad_plot} represents unstable equilibria, where the ground state energy is higher than the equilibrium on the right branch with the same number of particles (and thus the same quantum numbers).

\begin{figure}[htc]
\begin{center}
 \includegraphics[width=6in]{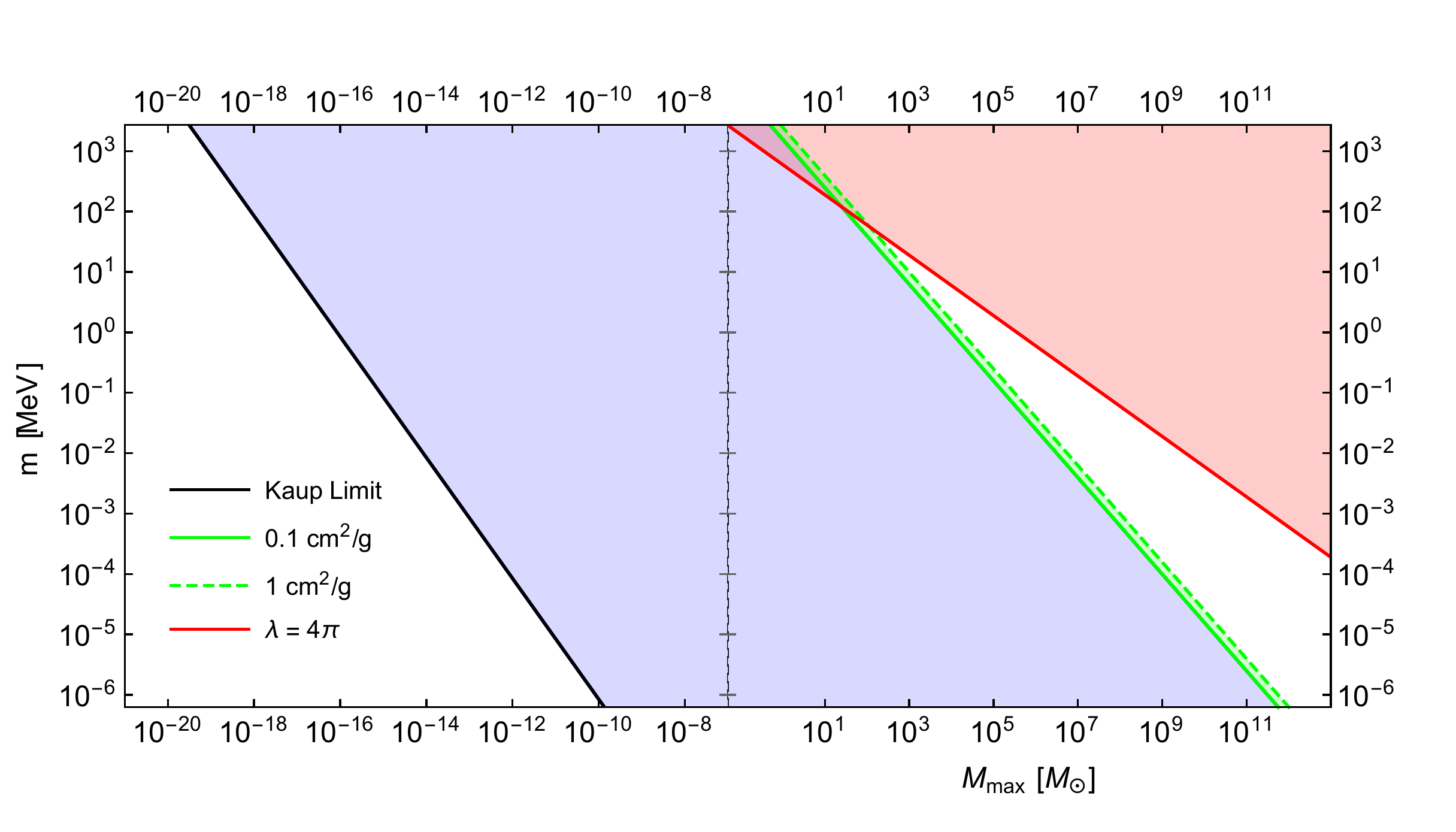}
\caption{The maximum mass of a boson star with  \emph{repulsive} self-interactions satisfying Eq. (\ref{bounds}), as a function of DM particle mass $m$. The green band is the region consistent with solving the small scale problems of collisionless cold DM. The blue region represents generic allowed interaction strengths (smaller than $0.1$ cm$^2$/g) extending down to the Kaup limit which is shown in black. The red shaded region corresponds to $\lambda\gtrsim4\pi$.  Note that the horizontal axis is measured in solar masses $M_\odot$.}
\label{M_Rep_plot}
\end{center}
\end{figure}

If we take the allowed range of $\lambda$ to be given by Eq. (\ref{bounds}), then we find the following range for $M_\text{max}^\text{rep}$:
\begin{equation}
 \Big(\frac{1 \text{ MeV}}{m}\Big)^{5/4} 3.42\times10^{4} M_{\odot} 
      \lesssim M_\text{max}^\text{rep}
      \lesssim \Big(\frac{1 \text{ MeV}}{m}\Big)^{5/4} 6.09\times10^{4} M_{\odot},
\end{equation}
where $M_{\odot}=1.99\times10^{30}$ kg is the solar mass.  The range of masses allowed by these inequalities are represented in Figure \ref{M_Rep_plot}.  Because of the strength of the repulsive interactions, these solutions can have masses several orders of magnitude above $M_\odot$.  If there is a significant number of such objects in the Milky Way,  it could have important observational signatures.  However, a detailed analysis of the formation of these objects is required, in order  to give some indication of whether DM boson stars in galaxies have masses close to the maximum value or lower. 

\subsection{Attractive Interactions}\label{3.2}
If DM self-interactions are attractive, then the method of \cite{Colpi:1986ye} does not apply.  However, assuming relativistic corrections are negligible, one can instead solve the nonrelativistic equations of motion numerically and analyze the solutions.  To be precise, the dynamics of a dilute, nonrotating BEC are governed by the Gross-Pit\"aevskii equation for a single condensate wavefunction $\phi(r,t)=\psi(r)e^{-iEt}$ \cite{Boehmer:2007um}
\begin{equation}\label{GP}
 E \psi(r) = \Big(-\frac{\vec{\nabla}^2}{2m} + V(r) + \frac{4\pi a}{m}|\psi(r)|^2 \Big)\psi(r)
\end{equation}
where $V$ is the trapping potential, which in our case is the gravitational potential of the BEC and satisfies the Poisson equation
\begin{equation} \label{Poisson}
\vec{\nabla}^2 V(r) = 4 \pi G m \rho(r). 
\end{equation}
The s-wave scattering length $a$ is related to a dimensionless $\phi^4$ coupling $\lambda$ by $a=\lambda/(32\pi m)$.

Here, $\rho(r) = m\cdot n(r) = m\cdot |\psi(r)|^2$ is the mass density of the condensate, which is normalized such that $\int{d^3 r \rho(r)} = M$, the total mass.  The three terms on the right-hand side of Eq. (\ref{GP}) correspond to the kinetic, gravitational, and self-interaction potentials, respectively.  As our notation signifies, we will assume that the density function is spherically symmetric, i.e. $\rho(\vec{r}) = \rho(r)$, which should be correct for a ground state solution.

\begin{figure}[htc]
\centering
\begin{minipage}{0.46\textwidth}
\centering
 \includegraphics[width=0.9\textwidth]{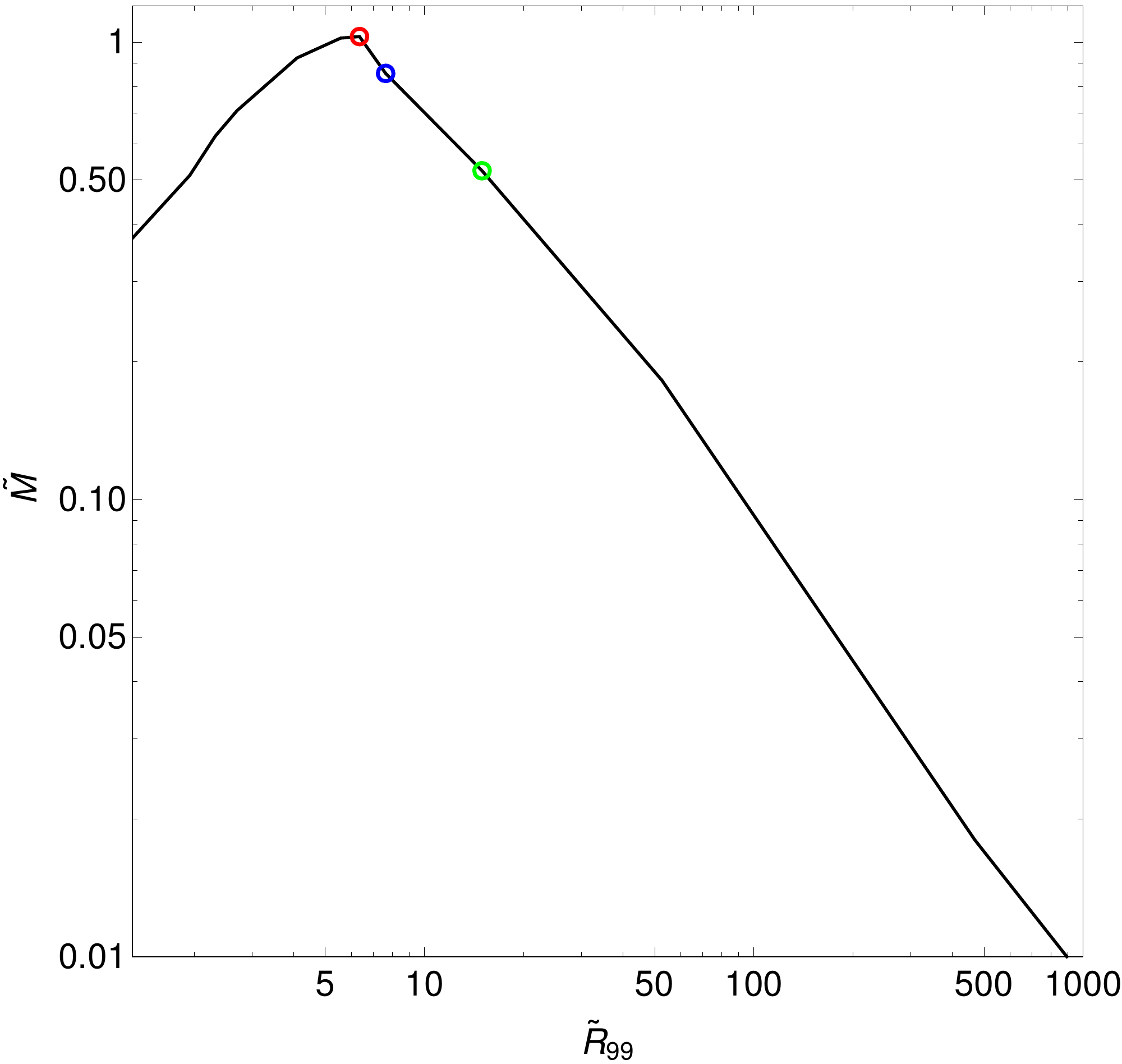}
\caption{The mass-radius relation for a boson star with attractive interactions. The three circles correspond to the density profiles in Figure \ref{Att_Density_plot}.  The dimensionless variables in the plot are defined in terms of the dimensionful ones as $\displaystyle{\tilde{M}=\sqrt{\frac{\lambda}{32\pi}}\frac{M}{M_P}}$ and $\displaystyle{\tilde{R}_{99}=\sqrt{\frac{32\pi}{\lambda}}\frac{m^2}{M_P}R_{99}}$.}
\label{Att_MassRad_plot}
\end{minipage}\hfill
\begin{minipage}{0.46\textwidth}
\centering
 \includegraphics[width=.9\textwidth]{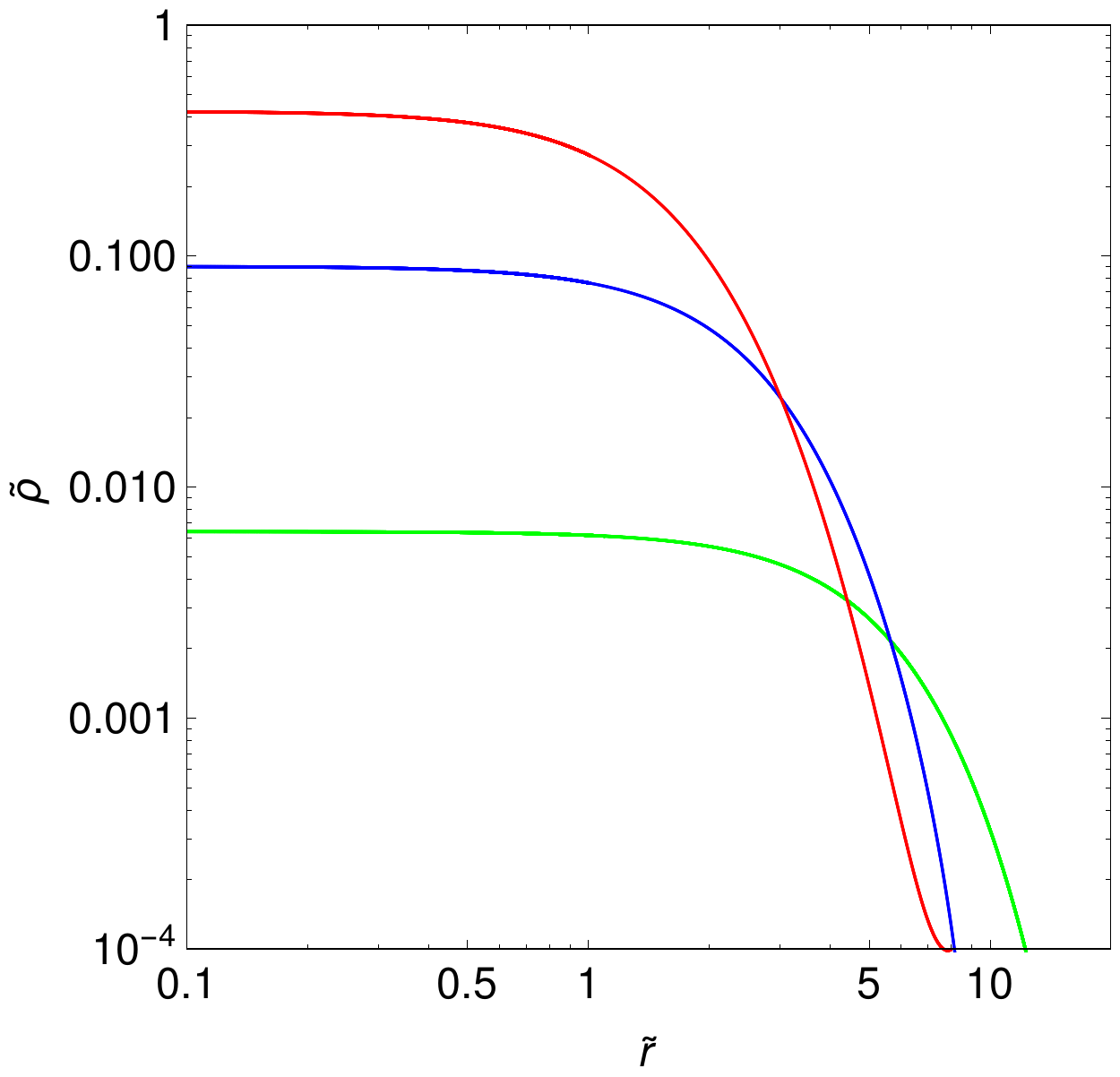}
\caption{Three examples of density profiles in the case of attractive interactions. The red profile corresponds to the profile of the maximum mass equilibrium, while the blue and green are taken on the stable branch of equilibria.  The dimensionless variables in the plot are defined in terms of the dimensionful ones as $\displaystyle{\tilde{\rho}=\frac{\lambda}{m^4}\rho}$ and $\displaystyle{\tilde{r}=\sqrt{\frac{32\pi}{\lambda}}\frac{m^2}{M_P}r}$.}
\label{Att_Density_plot}
\end{minipage}
\end{figure}

Because the Gross-Pit\"aevskii + Poisson system (hereafter GP, defined by Eqs. (\ref{GP}) + (\ref{Poisson})) cannot be solved analytically in general, we use a shooting method to integrate the system numerically over a large range of parameters.  As boundary conditions, we choose the values of $\psi(0)$ and $V(0)$ so that both functions are regular as $r\rightarrow 0$, and so that asymptotically $\psi(r)\rightarrow 0$ and $r V(r)\rightarrow 0$ exponentially as $r\rightarrow\infty$.  Some examples of integrated density functions are given in Figure \ref{Att_Density_plot}.  Our numerical procedure requires the following rescaling of the dimensionful quantities:
\begin{align}
 \psi &= \sqrt{\frac{m}{4\pi G}}\frac{1}{|\tilde a|} \tilde\psi \qquad
 V - E = \frac{m}{|\tilde a|} \tilde V \nonumber \\
 a &= m G |\tilde a| \qquad \qquad \quad
 r = \frac{\sqrt{|\tilde a|}}{m} \tilde r,
\end{align}
where the dimensionless quantities on the RHS are denoted with a tilde.  The equations take the form
\begin{align}
 \left[-\frac{1}{2}\tilde{\nabla}^2 + \tilde{V} 
	    - |\tilde{\psi}^2|\right]\tilde{\psi} = 0 \nonumber \\
 \tilde{\nabla}^2\tilde{V} = |\tilde{\psi}^2|,
\end{align}
where $\tilde{\nabla}$ denotes a gradient with respect to $\tilde{r}$, and we have explicitly taken $a<0$.  These are the equations we solve. Similar rescaled equations were used in \cite{Guzman1}, but for repulsive interactions, and unlike \cite{Guzman1}, we also scale away the scattering length $a$.  This makes  our solutions valid for any generic $a<0$.

In Figure \ref{Att_MassRad_plot} we show the mass-radius relation for the bosonic stars, which agrees well with the results obtained in \cite{Chavanis:2011-2}.  As in the repulsive case, there is a maximum mass for these condensates, but this mass is significantly smaller for attractive interactions.  For parameters satisfying Eq. (\ref{bounds}), our analysis shows that condensates of this type would be light and very dilute, having masses $<1$ kg and radii $R\sim\mathcal{O}$(km).  (Our assumption that the General Relativistic effects could be neglected in this case is therefore well supported a posteriori.)

One can arrive at a good, order of magnitude analytic estimate on the size and mass of condensates by a variational method which minimizes the total energy.  To this end, we follow the approach of \cite{Chavanis:2011zi} by using the GP energy functional,
\begin{equation}\label{Energy}
 E[\psi] = \int{d^3r\Bigg[\frac{|\vec{\nabla}\psi|^2}{2m} + V |\psi|^2} + \frac{2\pi a}{m}|\psi|^4\Bigg].
\end{equation}
As input, we choose an ansatz for the wavefunction $\psi(r)$, and subsequently compute the energy of the condensate by integrating Eq.~(\ref{Energy}) up to some maximum size $R$.  Minimizing the energy with respect to $R$ should give a good estimate for the size of stable structures.  Note that the gravitational potential $V(r)$ must be chosen self-consistently to satisfy Eq. (\ref{Poisson}) for a given choice of $\psi(r)$.

In order to illustrate the salient features of the method, we will choose a simple ansatz for the wavefunction:
\begin{equation} \label{Wavefunction}
 \psi(r) = 
 \begin{cases}\sqrt{\frac{3N}{4\pi R^3}}e^{i r/R} &\text{if $r \leq R$,}\\
    0 &\text{if $r>R$,}
\end{cases}
\end{equation}
which is normalized as above.  Performing the energy integral gives the result
\begin{equation}
 E = N \Big[\frac{A}{R^2} - \frac{B N}{R} + \frac{3A N a}{R^3}\Big],
\end{equation}
where $A\equiv 1/(2m)$ and $B\equiv 6 G m^2/5$.  Minimizing $E(R)$ with respect to $R$ gives two critical points
\begin{equation}
 R_\text{c} = \frac{A}{B N}\Big(1\pm \sqrt{1+\frac{9a}{A/B}N^2}\Big).
\end{equation}
In this calculation, a natural length scale $X\equiv A/B$ emerges.  For any $a\neq 0$ (repulsive or attractive), the minimum of the energy lies at the solution with the ``+'' sign, i.e.
\begin{equation} \label{Radius}
 R_0 = \frac{X}{N}\Big(1 + \sqrt{1+\frac{9a}{X}N^2}\Big).
\end{equation}

In the case of attractive interactions, there is a critical number of particles $N_\text{max} \equiv \sqrt{X/(9|a|)}$, above which the real energy minimum disappears and no stable condensate exists.  Using $M_\text{max} = m N_\text{max}$, this analysis sets a value for the maximum mass for stable condensates with attractive interactions:
\begin{equation} \label{M_Att}
 M < M_\text{max}^\text{att} = m \sqrt{\frac{X}{9|a|}} = \sqrt{\frac{320}{27}}\frac{M_\text{P}}{\sqrt{|\lambda|}}.
\end{equation}
The corresponding limit on the radius is a lower bound, attractive boson stars being stable only for
\begin{equation} \label{R_Att}
 R > R_\text{min}^\text{att} = \sqrt{\frac{15}{16}|\lambda|}\frac{M_\text{P}}{m^2}.
\end{equation}
Note that while the coefficient depends on the details of the wavefunction ansatz, the scaling relations $M_\text{max}^\text{att}\sim M_\text{P}/\sqrt{|\lambda|}$ and $R_\text{min}^\text{att}\sim \sqrt{\lambda}M_\text{P}/m^2$ are completely generic.

\begin{figure}[htc]
\begin{center}
 \includegraphics[width=6in]{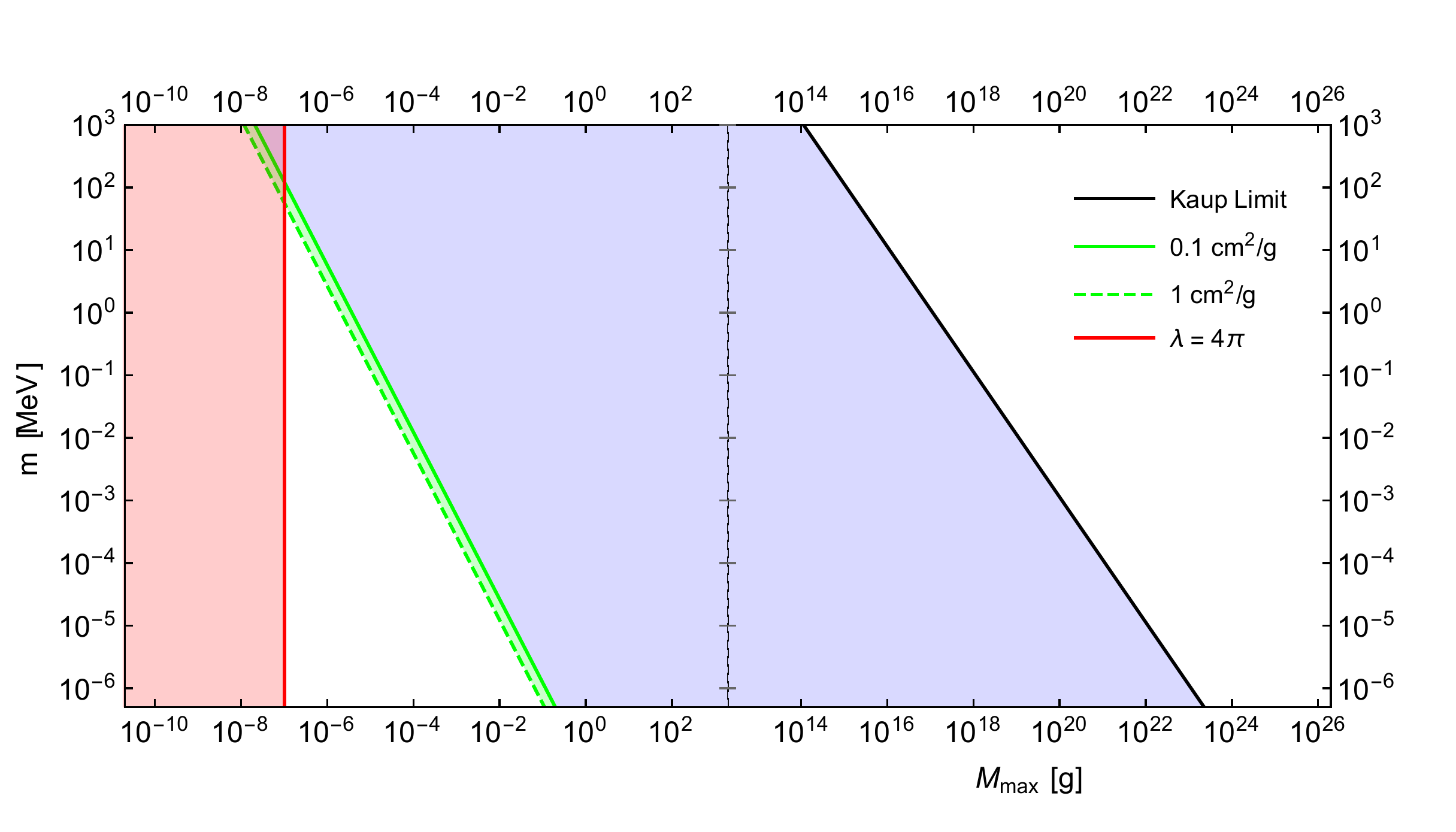}
\caption{The maximum mass of a boson star with  \emph{attractive} self-interactions satisfying Eq.~(\ref{bounds}), as a function of DM particle mass $m$. The green band is the region consistent with solving the small scale problems of collisionless cold DM. The blue region represents generic allowed interaction strengths (smaller than $0.1$ cm$^2$/g) extending up to the Kaup limit which is shown in black. The red shaded region corresponds to $\lambda\gtrsim4\pi$.  Note that the horizontal axis is measured in grams.}
\label{MaxMass_Att}
\end{center}
\end{figure}

Using Eq. (\ref{bounds}), we find
\begin{equation}
 \Big(\frac{1 \text{ MeV}}{m}\Big)^{3/4} 7.37\times10^{-9} \text{ kg} 
	\lesssim M_\text{max}^\text{att}   
	\lesssim \Big(\frac{1 \text{ MeV}}{m}\Big)^{3/4} 1.31\times10^{-8} \text{ kg}
\end{equation}
The range of masses allowed by these inequalities is given by the green band in Figure \ref{MaxMass_Att}.  We plot the maximum masses over many orders of magnitude, between $1$ eV and $1$ GeV, but the maximum mass of boson stars with such strong attractive self-interactions is still $<1$ kg.

Note that the numerical results agree well with the predictions of the variational method to within an order of magnitude, even for the na\"ive constant density ansatz in Eq. (\ref{Wavefunction}).  These estimates can be improved further by a more robust ansatz for the wavefunction.

As an example of a physical model, field theories describing axions exhibit an attractive self-coupling through the expansion of the axion potential $V(A)=m^2 f^2\Big(1-\cos(A/f)\Big)$, where $A$ is the axion field, $m$ is the axion mass, and $f$ is the axion decay constant.  Gravitationally bound states, particularly in the context of QCD axions, have become the topic of much recent interest \cite{Barranco_2011,Eby_2015,Guth_2015}.  These states typically have maximum masses of roughly $10^{-11} M_\odot$, far below the bounds set in this section, because the couplings are typically many orders of magnitude smaller.

As we pointed out in the introduction, in the case of attractive interactions the potential  is unbounded from below since $\lambda < 0$. Therefore there must exist higher dimensional operators suppressed by some cutoff. The first irrelevant operator with a $\mathbb{Z}_2$ symmetry is $\phi^6/\mu_\text{c}^2$ where $\mu_\text{c}$ is the cutoff scale. We will now set a lower limit for $\mu_\text{c}$ by requiring that the $\phi^6$ term is negligible with respect to the $\phi^4$ term for typical boson star field values. Assuming that the kinetic energy of the field is negligible, the energy density is roughly equal to the potential. The maximum mass and minimum radius in Eqs. (\ref{M_Att}) and (\ref{R_Att}) can also be used to estimate the energy density as $\rho \approx M_\text{max}/R_\text{min}^3 \approx m^6/|\lambda|^2 M_\text{P}^2$. Now we can estimate the field value $\tilde{\phi}$ inside the boson star with attractive interactions to be
\begin{equation}
|\tilde{\phi}| \approx \frac{m}{\sqrt{2|\lambda|}} \left(1+ \left(1-\frac{4m^2}{ |\lambda| M_\text{P}^2}\right)^{\tfrac{1}{2}}\right)^{\tfrac{1}{2}} \approx \frac{m}{\sqrt{|\lambda|}}.
\end{equation}
Requiring $|\lambda|\tilde{\phi}^4 \gg \tilde{\phi}^6/\mu_\text{c}^2$ we obtain the inequality $\mu_\text{c} \gg m/ |\lambda|$.


\section{Conclusions}


In this paper we studied the possibility that self-interacting bosonic DM forms stars. We  assumed that self-interactions are mediated by a $\lambda \phi^4$ interaction and we investigated what type of stars can be formed in the case of both attractive and repulsive self-interactions, giving particular emphasis to the parameter phase space of masses and couplings where the DM bosons alleviate the problems of collisionless DM. We have considered DM particles that populate the BEC ground state. We estimated the maximum mass where these dark stars are stable, the mass-radius relation and the density profile for generic values of DM mass and self-interacting coupling $\lambda$. 

We leave several things for future work. The first and most important is the mechanism of formation for these bosonic dark stars.  Sufficiently strong self-interactions can lead to the gravothermal collapse of part or the whole amount of DM to dark stars~\cite{Balberg:2002ue}. In this case, DM self-interactions can facilitate the formation of bosonic stars because DM particles get confined to deeper self-gravitating wells simply by expelling high energetic DM particles out of the core. As the core loses energy, the virial theorem dictates that the core shrinks and heats up the same time. This leads to further energy loss and thus to the gravothermal collapse. Such a scenario could also explain why the  black hole at the center of the Galaxy is so heavy, since DM bosonic stars could provide the initial seed required for the further growth of the supermassive black hole~\cite{Pollack:2014rja}. It is interesting to note that boson stars can coexist in equilibrium with black holes, as shown in \cite{Herdeiro1,Herdeiro2}. One should also notice that if the whole density of DM collapses to dark stars, one does not have to be within the narrow band of parameter space depicted in Figures \ref{M_Rep_plot} and \ref{MaxMass_Att}. Another possibility is the creation of high DM density  regions due to adiabatic contraction, caused by baryons~\cite{Blumenthal:1985qy,Gustafsson:2006gr}. Moreover, bosonic DM particles can get trapped inside regular stars via DM-nucleon collisions. The DM population is inherited by subsequent white dwarfs that, in case of supernovae 1a explosions, can leave the bosonic matter intact, either alone or with some baryonic matter~\cite{Kouvaris:2010vv}. 

Asymmetric bosonic dark stars where no substantial number of annihilations take place will not be very visible in the sky, although present. Gravitational lensing could be one way to deduce the presence of such stars in the universe. Additionally, if the DM boson interacts with the Standard Model particles via some portal (e.g. kinetic mixing between a photon and a dark photon),  thermal Bremmstrahlung could potentially produce   an observable amount of luminosity. This is particularly interesting since such a photon spectrum would probe directly the density profile of the boson star. Bosonic stars could also disguise themselves as ``odd" neutron stars. For example, it is hard to explain sub-millisecond pulsars with typical neutron stars. XTE J1739-285 could possibly be such a case, since it allegedly rotates with a frequency of 1122Hz \cite{Kaaret:2006gr}. Compact enough bosonic stars would have no problem to explain such high rotational frequencies. Another possibility is the observation of compact stars with masses higher than the maximum mass a neutron star can support. Such might be the case of the so-called ``black widow" PSR B1957+20, with a mass of 2.4 solar masses~\cite{vanKerkwijk:2010mt}.
Therefore, abnormal neutron stars can well be the smoking gun for the existence of asymmetric dark stars either with fermionic constituents like~\cite{Kouvaris:2015rea}, or with the bosonic  ones studied here.

{\bf Acknowledgements}\\
The research of C.K. and N.G.N. is supported by the Danish National Research Foundation, Grant No. DNRF90. J.E. is supported by a Mary J. Hanna Fellowship. L.C.R.W.'s research is partially supported by a faculty development award at UC. C.K. and  L.C.R.W. acknowledge support by the Aspen Center for Physics through the  NSF Grant PHY-1066293. J.E. and L.C.R.W. also acknowledge M. Ma, C. Prescod-Weinstein, and P. Suranyi for valuable discussions about Bose-Einstein Condensation.


\end{document}